\documentclass[12pt]{article}
\usepackage{amsbsy}
\usepackage{amsfonts}
\usepackage{amsmath}
\usepackage{amssymb}
\usepackage{apacite}
\usepackage{setspace}
\usepackage{color,soul}
\usepackage{graphicx}
\usepackage{bm,multicol}
\usepackage[english]{babel}
\usepackage{natbib}
\usepackage[T1]{fontenc}
\usepackage[utf8]{inputenc}
\usepackage{authblk}
\usepackage{xcolor}

\newcommand{\newc}{\newcommand}
\newc{\N}{\mbox{N}}
\newc{\1}{\bf{1}}

\begin{document}
\title{BIC extensions for order-constrained model selection}

\author{}
\author{J. Mulder \& A. E. Raftery}
\date{}
\maketitle

\begin{abstract}
The Schwarz or Bayesian information criterion (BIC) is one of the most widely used tools for model comparison in social science research. The BIC however is not suitable for evaluating models with order constraints on the parameters of interest. This paper explores two extensions of the BIC for evaluating order constrained models, one where a truncated unit information prior is used under the order-constrained model, and the other where a truncated local unit information prior is used. The first prior is centered around the maximum likelihood estimate and the latter prior is centered around a null value. Several analyses show that the order-constrained BIC based on the local unit information prior better functions as an Occam's razor for evaluating order-constrained models and results in lower error probabilities. The methodology based on the local unit information prior is implemented in the R package `BICpack' which allows researchers to easily apply the method for order-constrained model selection. The usefulness of the methodology is illustrated using data from the European Values Study.
\end{abstract}

\section{Introduction}
The Bayesian information criterion (BIC) is one of the most commonly used model evaluation criteria in social research, for example for categorical data \citep{Raftery:1986}, event history analysis \citep{Vermunt}, or structural equation modeling \citep{Raftery1993,Lee:2007}. The BIC, originally proposed by \cite{Schwarz:1978}, can be viewed as a large sample approximation of the marginal likelihood \citep{Jeffreys} based on a so-called unit information prior. This unit information prior contains the same amount of information as would a typical single observation \citep{Raftery:1995}.

The BIC has several useful properties. First, it can be used as a default quantification of the relative evidence in the data between two statistical models. Second, it can straightforwardly be used for evaluating multiple statistical models simultaneously. Third, it is consistent for most well-behaved problems in the sense that the evidence for the true model converges to infinity \citep{KassWass1995}. Fourth it behaves as an Occam's razor by balancing model fit (quantified by the log likelihood function at the maximum likelihood estimate (MLE)) and model complexity (quantified by the number of free parameters). Fifth, it is easy to compute using standard statistical software: only the MLEs, the maximized loglikelihood, the sample size, and the number of model parameters are needed to compute it. All these useful properties have contributed to the popularity and usefulness of the BIC in social research.

Despite the general applicability of the BIC, it is not suitable for evaluating statistical models with order constraints on certain parameters. In a regression model for instance it may be expected that the first predictor has a larger effect on the outcome variable than the second predictor, and the second predictor is expected to have a larger effect than the third predictor. This can be translated to the following order-constrained model, $M_1:\beta_1>\beta_2>\beta_3$, where $\beta_k$ denotes the effect of the $k$-th predictor on the outcome variable. This model can then be tested against conflicting models, such as a model with competing order constraints, e.g., $M_2:\beta_3>\beta_1>\beta_2$, a model where the effects are expected to be equal, $M_3:\beta_1=\beta_2=\beta_3$, or the complement of these models, denoted by $M_4$. Under the complement model $M_4$, the true values for the $\beta$'s do not satisfy any of the constraints under models $M_1$, $M_2$, or $M_3$. The reason that the BIC is not suitable for testing models with order constraints is that the number of free parameters does not properly capture the complexity of a model. In the above model $M_1$, all three $\beta$ parameters are free parameters but saying that $M_1$ is equally complex as a model with no constraints, i.e., $(\beta_1,\beta_2,\beta_3)\in\mathbb{R}^3$, seems incorrect. Furthermore, the BIC is based on the Laplace approximation of the marginal likelihood. It is as yet unclear how well the approximation performs in the case of models with order constraints. {The complicating factor is that the approximation assumes that the maximum value of the integrand is an interior point of the integrated region. This assumption is violated if the maximum likelihood (or posterior mode) does not lie in the integrated region.}

Testing order constraints is particularly useful because effect sizes can only be interpreted relative to each other in the study, and relative to the field of research \citep{Cohen:1988}. An effect size of, say, 0.3, of educational level on attitude towards immigrants may seem substantial for a sociologist, while 0.3 may not be interesting when it quantifies the effect of a medical treatment on the amount of pain of a patient. Thus, instead of interpreting the magnitude of an effect by its estimated value it is more informative to interpret them relative to each other, as is done using order-constrained model selection. This way we are able to determine which effects dominate other effects in the study. Furthermore, order-constrained model selection will be useful for testing scientific expectations which can often be formulated using order constraints (examples will be given in Section 2, but also see \cite{Klugkist:2005}, \cite{Hoijtink:2011}, \cite{Braeken:2015}, \cite{MulderPericchi:2018}, \cite{MulderFox:2018}, for example). By testing order-constrained models we are able to quantify the relative evidence in the data for a scientific theory against competing theories. 

As will be shown, order-constrained models are also naturally specified when one is interested in the effect of an ordinal categorical variable on an outcome variable of interest. Finally, the inclusion of order constraints results in more statistical power. This can be explained by the smaller subspace for the parameters under an order-constrained model in comparison to an equivalent model without the order constraints. The order constraints make the model `less complex' resulting in a smaller penalty for model complexity, and thus in more evidence for an order-constrained model that is supported by the data.

{In this paper we explore how the BIC can be extended to enable order-constrained model selection. First a unit information prior is considered that is truncated in the order-constrained subspace. This results in a BIC that may not properly incorporate the relative complexity of an order-constrained model. For this reason an alternative local unit information prior is considered which is centered around a null value. This prior results in a BIC that properly incorporates the relative fit and complexity of order-constrained models.} To ensure general utilization of the order-constrained BIC based on the local unit information, the R package `{\tt BICpack}' has been developed for order-constrained model selection in popular models such as generalized linear models, survival models, and ordinal regression models.

To our knowledge there have been two other proposals for the BIC for evaluating models with order (or inequality) constraints by \cite{Romeijn:2012} and \cite{Morey:2014}, and we will compare our proposal to theirs. 

Before presenting the new methodology we first motivate the importance and usefulness of evaluating statistical models with order constraints on the parameters of interest in the context of the European Values Study in Section 2. Subsequently, BIC approximations of the marginal likelihood under an order-constrained model is discussed in Section 3. Section 4 provides a numerical evaluation of the methods. Section 5 explains how to apply the new method for testing social theories in the European Values Study. We end the paper with some concluding remarks in Section 6.

\section{Order-constrained model selection in social research}
In this section we present two situations where order-constrained model selection is useful. First, theories often make an assumption about the relative importance of certain predictors on an outcome variable. This can be formalized by specifying order constraints on the effects of these predictor variables. We will show this in Application 1 using the Ethnic Competition Theory \citep{Scheepers:2002}. Second, a researcher may have an expectation about the direction of an effect of a predictor variable with an ordinal measurement level. When modeling this ordinal predictor variable using dummy variables, the expected directional effect can be translated to a set of order constraints on the effects of these dummy variables. This will be shown in Application 2 by considering Inglehart's Generational Replacement Theory.

\subsection{Application 1: Assessing the importance of different dimensions of socioeconomic status}
In most European countries, the majority of immigrants are located in the lower strata of society. For this reason lower-strata members of the European majority population who hold similar social positions as the ethnic minorities, having a relatively low social class, low educational level, or low income level, will on average compete more with ethnic minorities than other citizens in the labour market. 
Therefore Ethnic Competition Theory \citep{Scheepers:2002} would predict that higher social class, educational level, or income level would result in a more positive attitude towards immigrants. Furthermore, it is likely that social class (which reflects the type of job a person has) has the largest impact because one's social class is directly related to the labour market. The effects of education is less direct and therefore it is expected that one's educational level has a lower impact on attitude towards immigrants than social class. Finally it would be expected that the effect of income would be the lowest, but still positive. This expectation will be formalized in model $M_1$ which is provided below.

Alternatively, due to the importance of education in shaping one's identity \citep{Cohen:2013,Waal:2015}, it might be expected that education is the most important factor explaining one's attitude towards immigrants, followed by social class and income for which no specific ordering is expected (formalized in model $M_2$). A third hypothesis is that all three dimensions have an equal and positive effect on attitudes towards immigrants (model $M_3$). Finally, it may be that neither of these three hypotheses are true (model $M_4$).

To evaluate these expectations we first write down the linear regression model where the attitude towards immigrants is the outcome variable, and social class, educational level, and income are the predictor variables while controlling for age. The $i$-th observation is modeled as follows,
\begin{eqnarray}
\label{model1}\text{attitude}(i) &=& \theta_0 ~+ ~\text{class}(i)\times \theta_{\text{class}} ~+ ~\text{education}(i)\times\theta_{\text{education}}\\
\nonumber &&+~ \text{income}(i)\times\theta_{\text{income}}~ + ~\text{gender}(i)\times\theta_{\text{gender}}+\text{error}(i),
\end{eqnarray}
for $i=1,\ldots,n$. The predictor variables are all standardized. In \eqref{model1}, $\theta_{\text{class}}$, $\theta_{\text{education}}$, and $\theta_{\text{income}}$ are the standardized effects for social class, educational level, and income, respectively, $\theta_{\text{gender}}$ is the standardized effect of gender, and the error is assumed to be normally distributed with unknown variance.

The four expectations given above can be formalized using competing statistical models with different order constraints on the standardized effects,
\begin{eqnarray}
\nonumber M_1&:&\theta_{\text{class}}>\theta_{\text{education}}>\theta_{\text{income}}>0\\
\label{App1} M_2&:&\theta_{\text{education}}>(\theta_{\text{class}},\theta_{\text{income}})>0\\
\nonumber M_3&:&\theta_{\text{class}}=\theta_{\text{education}}=\theta_{\text{income}}>0\\
\nonumber M_4&:&\text{``neither $M_1$, $M_2$, nor $M_3$''}
\end{eqnarray}
where $\theta_{\text{class}}$, $\theta_{\text{education}}$, and $\theta_{\text{income}}$ denote the effects of social class, educational level, and income on attitude towards immigrants, respectively. Consequently the goal is to quantify the evidence in the data for these three models to determine which model receives most support. 

Note that nuisance parameters (e.g., effects of control variables) are omitted in the above formulation of the models of interest to simplify the notation. Further note that additional competing constrained models could be formulated in this context as well. For the current application however we restrict ourselves to these models.

\subsection{Application 2: The importance of postmaterialism for young, middle and old generations}
Experiences in pre-adult years are known to have a crucial impact on the development of basic values in later life. Due to the increase in welfare in recent decades, Generational Replacement Theory predicts that values have shifted from older generations to newer generations. In particular postmaterialistic values, such as the desire for freedom, self-expression, and quality of life, are expected to have increased for younger generations as a result of improved economic standards in western countries \citep{Inglehart:1999,WelzelInglehart:2005}. 

In the European Values Study, generation was operationalized using an ordinal variable with three categories corresponding to a young, middle or old generation. Similarly, postmaterialism has been measured on an ordinal scale as well, having three categories. When setting the younger generation as the reference group and using dummy variables for the middle and older generations, the Generational Replacement Theory can be translated to an order-constrained model ($M_1$). We contrast it with a model that assumes no generation effect on postmaterialism ($M_0$) and with a complementary model that assumes neither an increased effect nor a zero effect ($M_2$). 

The models of interest can be summarized as follows
\begin{eqnarray}
\nonumber M_0&:&\theta_{\text{old}}=\theta_{\text{middle}}=0\\
\label{App2} M_1&:&\theta_{\text{old}}<\theta_{\text{middle}}<0\\
\nonumber M_2&:&\text{``neither $M_0$, nor $M_1$,''}
\end{eqnarray}

Furthermore we hypothesize that the inclusion of order constraints on the generational effects of interest results in an increase of power in comparison to testing the classical alternative, say, $M_3:\theta_{\text{young}}\not =\theta_{\text{middle}} \not = 0$ versus the null model $M_2:\theta_{\text{young}}=\theta_{\text{middle}}=0$. In terms of the BIC this implies we obtain more evidence against $M_0$ when testing it against the order-constrained model $M_1$ (if the constraints are supported by the data) than when testing $M_0$ against the unconstrained alternative $M_3$.

\section{BIC approximations of the marginal likelihood}
In this section, extensions of the BIC are derived for a model with order (or inequality) constraints on certain model parameters. Consider an order-constrained model $M_1$ with $d$ unknown model parameters, denoted by $\bm\theta$, which are restricted by $r_1$ order constraints, i.e., $M_1:\textbf{R}_1\bm\theta>\textbf{r}_1$ where $[\textbf{R}_1|\textbf{r}_1]$ is an augmented $r_1\times (d+1)$ matrix containing the coefficients of the order constraints under $M_1$. For example, order-constrained model $M_1:\theta_{\text{class}}>\theta_{\text{education}}>\theta_{\text{income}}>0$ in \eqref{App1} can be translated to 
\begin{equation}
\textbf{R}_1\bm\theta>\textbf{r}_1 \Leftrightarrow
\left[\begin{array}{cccccc}
0 & 1 & -1 & 0 & 0 & 0\\
0 & 0 & 1 & -1 & 0 & 0\\
0 & 0 & 0 & 1 & 0 & 0\\
\end{array}\right]
\left[\begin{array}{l}
\theta_{0}\\ \theta_{\text{class}}\\ \theta_{\text{education}}\\
\theta_{\text{income}}\\ \theta_{\text{gender}}\\ \sigma^2
\end{array}\right]>\left[\begin{array}{c}
0\\
0\\
0
\end{array}\right] ,
\label{Rr1}
\end{equation}
where the first element of $\bm\theta$ denotes the intercept, the fifth element the gender effect, and the sixth element denotes the error variance, which are nuisance parameters. The order-constrained model is nested in an unconstrained model which will be denoted by $M_u$. The likelihood function under $M_1$ is a truncation of the likelihood under an unconstrained model, i.e., $p_{1}(\textbf{D}|\bm\theta)=p(\textbf{D}|\bm\theta)\times I_{\bm\Theta_1}(\bm\theta)$, where $p(\textbf{D}|\bm\theta)$ denotes the likelihood function of the data $\textbf{D}$ under the unconstrained parameter space $\bm\Theta$. The prior for $\bm\theta$ under $M_1$ will be denoted by $p_1(\bm\theta)$. Two different types of priors will be considered for approximating the marginal likelihoods under $M_1$ and $M_u$.

\subsection{Truncated unit information prior}
First we assume that the unconstrained posterior mode, denoted by $\tilde{\bm\theta}_u$, falls in the inequality-constrained space of model $M_1$, i.e., $\textbf{R}_1\tilde{\bm\theta}>\textbf{r}_1$. The BIC approximation of the marginal likelihood under the inequality-constrained model is then obtained using a second-order Taylor expansion of the logarithm of the integrand around the posterior mode. This approximation introduces in an error that is $\mathcal{O}(n^{-1}).$\footnote{A quantity being $\mathcal{O}(n^{-1})$ implies that $n\mathcal{O}(n^{-1})$ converges to a constant, as $n\rightarrow \infty$.} Let us define $g(\bm\theta)=\log p_1(\textbf{D}|\bm\theta) + \log p_1(\bm\theta)$. Then, the marginal likelihood can be derived by
\begin{eqnarray*}
\log p_1(\textbf{D}) &=&\log \int_{\textbf{R}_1\bm\theta>\textbf{r}_1} p_1(\textbf{D}|\bm\theta)p_1(\bm\theta)d\bm\theta\\
&=& \log \int_{\textbf{R}_1\bm\theta>\textbf{r}_1} \exp\left\{g(\bm\theta)\right\}d\bm\theta\\
&=& \log \int_{\textbf{R}_1\bm\theta>\textbf{r}_1} \exp\left\{g(\tilde{\bm\theta}_u)+\frac{1}{2}(\bm\theta-\tilde{\bm\theta}_u)'\textbf{H}(\tilde{\bm\theta}_u)(\bm\theta-\tilde{\bm\theta}_u)\right\}d\bm\theta + \mathcal{O}(n^{-1})\\
&=& \log p_1(\textbf{D}|\tilde{\bm\theta}_u)+ \log p_1(\tilde{\bm\theta}_u) + \\
&&
\log \int_{\textbf{R}_1\bm\theta>\textbf{r}_1} \exp\left\{\frac{1}{2}(\bm\theta-\tilde{\bm\theta}_u)'\textbf{H}(\tilde{\bm\theta}_u)(\bm\theta-\tilde{\bm\theta}_u)\right\}d\bm\theta
+ \mathcal{O}(n^{-1})\\
&=& \log p_1(\textbf{D}|\tilde{\bm\theta}_u)+ \log p_1(\tilde{\bm\theta}_u)
+ \tfrac{d}{2}\log(2\pi)-\tfrac{1}{2}\log |-\textbf{H}(\tilde{\bm\theta}_u)|\\
&&+ 
\log \text{Pr} (\textbf{R}_1\bm\theta >\textbf{r}_1 |\textbf{D}, M_u)
+ \mathcal{O}(n^{-1}),
\end{eqnarray*}
where $\textbf{H}(\tilde{\bm\theta}_u)$ denotes the Hessian matrix of second-order partial derivatives of $g(\bm\theta)$ evaluated at $\tilde{\bm\theta}_u$. 

Hence, the only difference with the original derivation is that the resulting approximation also includes the posterior probability that the order constraints of $M_1$ hold under the larger unconstrained model $M_u$. From large sample theory, the unconstrained posterior mode can be approximated with the unconstrained maximum likelihood estimate (MLE), i.e., $\tilde{\bm\theta}_u\approx\hat{\bm\theta}_u$, and $-\textbf{H}(\tilde{\bm\theta}_u)\approx n I_E(\hat{\bm\theta}_u)$, where $I_E(\hat{\bm\theta}_u)$ is the expected Fisher information matrix of one observation (which can be obtained using standard statistical software). This introduces an additional approximation error of $\mathcal{O}(n^{-\frac{1}{2}})$. Subsequently the approximated logarithm of the marginal likelihood is given by
\begin{eqnarray}
\nonumber \log p_1(\textbf{D}) &=&\log p_1(\textbf{D}|\hat{\bm\theta}_u)+ \log p_1(\hat{\bm\theta}_u) + \tfrac{d}{2}\log(2\pi)
-\tfrac{d}{2}\log (n)-\tfrac{1}{2}\log |I_E(\hat{\bm\theta}_u)|\\
&&+\log \text{Pr} (\textbf{R}_1\bm\theta >\textbf{r}_1 |\textbf{D}, M_u) +\mathcal{O}(n^{-\frac{1}{2}}).
\label{approx1}
\end{eqnarray}
As was pointed out by \cite{Raftery:1995}, certain terms cancel out when plugging in the so-called unit information prior \cite[see also][]{KassWass1995}. The unit information prior has a multivariate normal distribution with mean equal to the MLE and variance equal to the inverse of the expected Fisher information matrix of one observation, i.e., $p^{UI}(\bm\theta)=N(\hat{\bm\theta}_u,I_E(\hat{\bm\theta}_u)^{-1})$. Under the constrained model $M_1$ we propose using a truncated unit information prior, i.e.,
\begin{equation}
\label{truncprior}
p^{UI}_1(\bm{\theta})=p^{UI}(\bm\theta)\times I(\textbf{R}_1\bm\theta > \textbf{r}_1)\times \text{Pr}^{UI}(\textbf{R}_1\bm\theta > \textbf{r}_1|M_u)^{-1},
\end{equation}
where the prior probability serves as a normalization contant so that the truncated unit information prior integrates to one, i.e.,
\begin{eqnarray*}
\text{Pr}^{UI}(\textbf{R}_1\bm\theta > \textbf{r}_1|M_u)
&=&\int_{\textbf{R}_1\bm\theta>\textbf{r}_1}
p^{UI}(\bm\theta)
d\bm\theta.
\end{eqnarray*}
{Evaluating the logarithm of the unconstrained unit information prior at the unconstrained MLE yields $\log p^{UI}(\hat{\bm\theta}_u)=-\frac{d}{2}\log(2\pi)+\frac{1}{2}\log |I_E(\hat{\bm\theta}_u)|$}, and therefore \eqref{approx1} becomes
\begin{eqnarray}
\nonumber \log p_1(\textbf{D}) &=&\log p_1(\textbf{D}|\hat{\bm\theta}_u)+ 
-\tfrac{d}{2}\log (n)+\log \text{Pr} (\textbf{R}_1\bm\theta >\textbf{r}_1 |\textbf{D}, M_u)\\
&& - \log \text{Pr}^{UI} (\textbf{R}_1\bm\theta >\textbf{r}_1 |M_u)
+ \mathcal{O}(n^{-\frac{1}{2}}).
\label{approx2}
\end{eqnarray}
The corresponding order-constrained BIC is then obtained by multiplying the logarithm of the approximated marginal likelihood by $-2$ and ignoring the error term. This yields
\begin{eqnarray}
\nonumber \text{OC-BIC}(M_1) &=& -2\log p_1(\textbf{D}|\hat{\bm\theta}_u)+ 
d\log (n)-2\log \text{Pr} (\textbf{R}_1\bm\theta >\textbf{r}_1 |\textbf{D}, M_u)\\
&& +2 \log \text{Pr}^{UI} (\textbf{R}_1\bm\theta >\textbf{r}_1 |M_u),
\label{BIC1}
\end{eqnarray}
where the first two terms form the ordinary BIC of model $M_1$ without the order constraints, and the additional third and fourth term are used for the evaluation of the order constraints of $M_1$ within $M_u$.

Next we consider the case where the unconstrained posterior mode does not lie in the inequality-constrained subspace of $M_1$, i.e., $\textbf{R}_1\tilde{\bm\theta} \not \ge\textbf{r}_1$. In this case the second-order Taylor expansion of $g(\bm\theta)$ around the unconstrained posterior mode (or MLE) may not be a good approximation. The rationale is that the mode under $M_1$, which will have a nonzero gradient, will lie on the boundary space where $\textbf{R}_1\bm\theta = \textbf{r}_1$.\footnote{In the case $\textbf{R}_1\tilde{\bm\theta} =\textbf{r}_1$, the second-order expansion may still be appropriate.} 

\begin{figure}[t]
\centering
\makebox{\includegraphics[width=13.0cm,height=6.0cm]{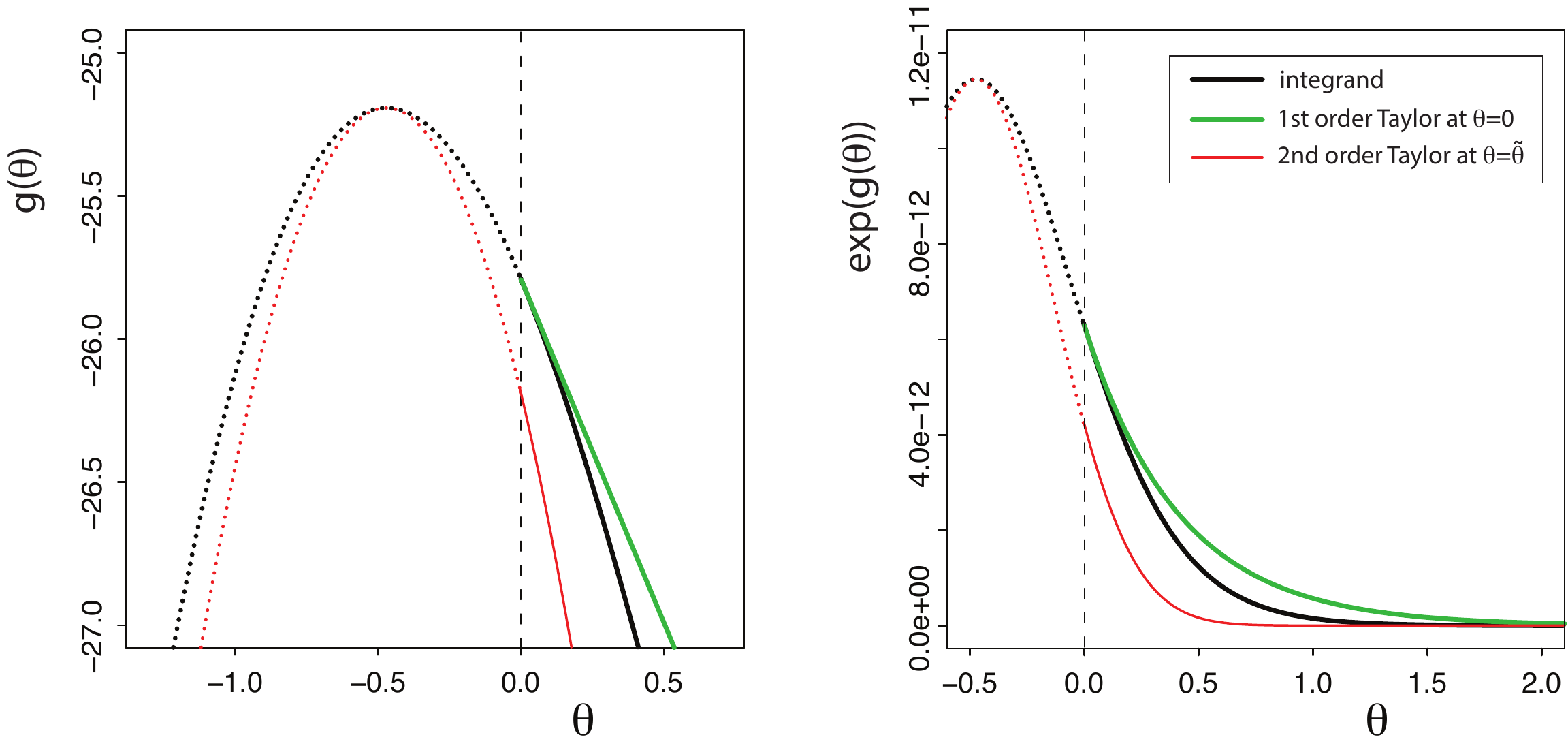}}
\caption{Plot of the log of prior times likelihood, $g(\theta)$, (black line), first-order Taylor approximation at $\theta=0$ (green line), and second-order Taylor approximation around the unconstrained posterior mode of $\tilde{\theta}\approx-.5$ (red line), for an inequality-constrained model $M_1:\theta\ge0$. The left panel displays the functions of the log scale and the right panel on the regular scale. The inequality-constrained region under $M_1:\theta\ge 0$ has solid lines, and the complement region has dotted lines.}
\label{figApprox}
\end{figure}

Because of the exponential tails of the normal distribution, a first-order Taylor expansion of $g(\bm\theta)$ at the posterior mode under $M_1$, denoted by $\tilde{\theta}_1$, seems more appropriate \citep{Avramidi:2000}. For example let us consider a simple inequality-constrained model, $M_1:\theta\ge 0$, and let the unconstrained mode be smaller than 0, i.e., $\tilde{\theta}<0$, so that the posterior mode under $M_1$ is located on the boundary, i.e., $\tilde{\theta}_1=0$, which has a negative gradient, $g'(0)<0$. The function $g(\theta)$ for such a situation is plotted in Figure \ref{figApprox} (black line, solid line under $\theta\ge 0$, dotted line under $\theta\not \ge 0$). The second-order Taylor approximation at the unconstrained mode is also plotted (red line; solid line under $\theta\ge 0$, dotted line under $\theta\not \ge 0$). 

A first-order Taylor expansion at $\tilde{\theta}_1=0$ can be used to approximate the function in the region $\theta \ge 0$ according to
\[
g(\theta) = g(0) + g'(0)\theta + \mathcal{O}(\theta^2).
\]
The marginal likelihood can then be approximated as follows
\begin{eqnarray*}
\log p_1(\textbf{D}) &=&\log \int_{\theta>0} p_1(\textbf{D}|\theta)p_1(\theta)d\theta
= \log \int_{\theta>0} \exp\left\{g(\theta)\right\}d\theta\\
&\approx& \log \int_{\theta>0} \exp\left\{g(0)+g'(0)\theta\right\}d\theta
= g(0) -\log(-g'(0)).
\end{eqnarray*}
Hence, instead of the normal distribution which is used to compute the integral in the case of a second-order Taylor expansion, an exponential distribution is used to compute the integral using this first-order Taylor expansion. The approximated line is also plotted in Figure \ref{figApprox} (green line).

The figure suggests that the second-order Taylor approximation at the unconstrained posterior mode is less accurate than the first-order Taylor approximation at the boundary point. This suggests that the approximated marginal likelihood under the inequality-constrained model will generally be better using first-order approximation at the boundary point in the case that the inequality constraints are not supported by the data. In the remainder of this paper however we shall use the second-order Taylor approximation at the unconstrained posterior mode both when the posterior mode does and does not lie in the subspace of the inequality-constrained model under investigation. 

When the order constraints are not supported by the data, the crudeness of the approximation is less important because the order-constrained model will not be selected because of the bad fit. Instead another, better fitting model will be selected for which the approximated marginal likelihood can be accurately estimated. Another reason for working with the second-order Taylor approximation is that it can easily be computed using \eqref{approx2}, also for more complex systems of inequality constraints on a multiple parameters, e.g., $\theta_1>\theta_2>\theta_3>0$, than when using a first-order Taylor approximation at boundary point of the inequality-constrained subspace where the mode is located. Numerical analyses presented later illustrate that the approximation error is acceptable when the posterior mode does not lie in the constrained subspace.

It has been argued that data-based priors, such as the unit information prior, may result in Bayes factors that do not function as an Occam's razor when evaluating inequality-constrained models \citep{Mulder:2014a,Mulder:2014b}. To see that this is also the case for the unit information prior, the approximated Bayes factor of an inequality-constrained model $M_1$ against an unconstrained model $M_u$ (where the inequality constraints are omitted) is given by
\[
B^{UI}_{1u} \approx \frac{\text{Pr} (\textbf{R}_1\bm\theta >\textbf{r}_1 |\textbf{D}, M_u)}
{\text{Pr}^{UI} (\textbf{R}_1\bm\theta >\textbf{r}_1 |M_u)}.
\]
This follows automatically from \eqref{approx2}. Now in the case of overwhelming evidence for $M_1$, i.e, $\textbf{R}_1\hat{\bm\theta}_u \gg \textbf{r}_1$, both the posterior probability and the prior probability based on the unit information prior will be approximately one, resulting in equal evidence for $M_1$ and $M_u$. This is a consequence of the fact that the unit information prior is concentrated around the MLE. Because both models fit the data equally well while the inequality-constrained model can be viewed as a less complex model (because a `smaller' subspace is spanned), this property suggests that the approximated Bayes factor does not properly function as an Occam's razor.

\subsection{Truncated local unit information prior}
Due to the behavior of the unit information prior when evaluating order-constrained models, we consider a `local' unit information prior with a mean that is located on the boundary of the inequality-constrained space \citep[we borrow the term `local' from][]{Johnson:2010}. Note that the boundary space is equal to the parameter space under the null model $M_0:\textbf{R}_1\bm\theta=\textbf{r}_1$. The rationale for centering the prior around the null space dates back to at least \cite{Jeffreys} who argued that when the null model is false the effects are expected to be close to the null; otherwise there is no point in testing the null. This implies that the prior under the alternative model should be located around the null value. {Furthermore there have been reports in the literature where the use of such local priors result in desirable selection behavior when evaluating order-constrained models }\citep[e.g.,][]{Mulder:2010,Mulder:2014a}. {Here we explore this class of priors for the BIC.}


We set the mean of the local unit information prior equal to the MLE under the null model, denoted by $\hat{\bm\theta}_0$. Furthermore, the covariance matrix will be equal to the covariance matrix of the unit information prior. Thus, the unconstrained local unit information prior can be written as $p^{LUI}(\bm\theta)=N(\hat{\bm\theta}_0,I_E(\hat{\bm\theta}_u)^{-1})$. The truncated prior under $M_1:\textbf{R}_1\bm\theta>\textbf{r}_1$ is then equal to
\begin{equation*}
\label{truncpriorL}
p^{LUI}_1(\bm{\theta})=p^{LUI}(\bm\theta)\times I(\textbf{R}_1\bm\theta>\textbf{r}_1)\times \frac{1}{\text{Pr}^{LUI}(\textbf{R}_1\bm\theta > \textbf{r}_1|M_u)}.
\end{equation*}

By applying formula (14) in \cite{Kass:1995}, changing the unit information prior to the local unit information prior, results in an approximated logarithm of the marginal likelihood of 
\begin{eqnarray}
\nonumber \log \hat{p}_1^{LUI}(\textbf{D}) &=&\log p(\textbf{D}|\hat{\bm\theta}_u)- \tfrac{d}{2}\log(n)-\tfrac{1}{2}(\hat{\bm\theta}_u-\hat{\bm\theta}_0)'I_E(\hat{\bm\theta})(\hat{\bm\theta}_u-\hat{\bm\theta}_0)\\
&&+\log \mbox{Pr}(\textbf{R}\bm\theta>\textbf{r}|\textbf{D},M_u)-\log\mbox{Pr}^{LUI}(\textbf{R}\bm\theta>\textbf{r}|M_u).
\label{BIC1-L}
\end{eqnarray}
Consequently, the approximated Bayes factor based on the local unit information prior of an inequality-constrained model against an unconstrained model is given by
\begin{equation}
B^{LUI}_{1u} \approx \frac{\text{Pr} (\textbf{R}_1\bm\theta >\textbf{r}_1 |\textbf{D}, M_u)}
{\text{Pr}^{LUI} (\textbf{R}_1\bm\theta >\textbf{r}_1 |M_u)}.
\label{BF1uLUI}
\end{equation}

Now in the case of overwhelming evidence for $M_1$, in the sense that $\textbf{R}_1\hat{\bm\theta}_u \gg \textbf{r}_1$, the Bayes factor will be equal to the reciprocal of the prior probability that the inequality constraints hold under the unconstrained local unit information prior, i.e., $B^{LUI}_{1u}\approx (\text{Pr}^{L} (\textbf{R}_1\bm\theta >\textbf{r}_1 |M_u))^{-1}$, which is strictly larger than one because the prior mean is located on the boundary of the constrained space where $\textbf{R}_1\bm\theta =\textbf{r}_1$. Note that this prior probability can be viewed as a quantification of the relative size of the inequality-constrained subspace. For example in the case of a diagonal covariance matrix, the prior probability of $k$ one-sided constraints, $\bm\theta>\textbf{0}$, is equal to ${2^{-k}}$, and the prior probability of $k$ order constraints, $\theta_1<\ldots<\theta_k$, is equal to $(k!)^{-1}$, similar to the Bayes factors proposed by \cite{Mulder:2010} and \cite{Morey:2014}.

Instead of working with \eqref{BIC1-L} we consider a slightly cruder approximation where the third term, which quantifies prior fit, is omitted. This yields
\begin{eqnarray}
\nonumber \log \hat{p}_1^{LUI*}(\textbf{D}) &=&\log p_u(\textbf{D}|\hat{\bm\theta}_u)- \tfrac{d}{2}\log(n)+\log \mbox{Pr}(\textbf{R}\bm\theta>\textbf{r}|\textbf{D},M_u)\\
&&-\log\mbox{Pr}^{LUI}(\textbf{R}\bm\theta>\textbf{r}|M_u).
\label{BIC1-L2}
\end{eqnarray}
The rationale for omitting this term is that we are not interested in quantifying prior misfit. Another reason is that expression \eqref{BIC1-L2} can be combined with the ordinary BIC approximation for an unconstrained model (i.e., `$\log p(\textbf{D}|\hat{\bm\theta}_u)- \tfrac{d}{2}\log(n)$') to obtain the approximated Bayes factor in \eqref{BF1uLUI}.

The final terms in \eqref{BIC1-L2} have the following intuitive interpretation. The first and second term can be interpreted as measures of model fit and model complexity of the unconstrained model where the inequality constraints are excluded (similar to the ordinary BIC approximation based on the unit information prior). The third term, which is the approximated posterior probability that the inequality constraints hold under the unconstrained model, can be interpreted as a measure of the relative fit of an order-constrained model $M_1$ relative to the unconstrained model $M_u$. Finally, the fourth term, which is the local prior probability that the order constraints hold under the unconstrained model, can be interpreted as a measure of the relative complexity of the order-constrained model $M_1$ relative to the unconstrained model $M_u$.

Thus \eqref{BIC1-L2} will behave as an Occam's razor when evaluating order-constrained models by balancing the fit and complexity of the order-constrained model. The corresponding order-constrained BIC based on the local unit-information prior then yields
\begin{eqnarray}
\nonumber \text{OC-BIC}^{*}(M_1) &=&-2\log p_u(\textbf{D}|\hat{\bm\theta}_u)+ d\log(n)-2\log \mbox{Pr}(\textbf{R}\bm\theta>\textbf{r}|\textbf{D},M_u)\\
&&+2\log\mbox{Pr}^{LUI}(\textbf{R}\bm\theta>\textbf{r}|M_u).
\label{BIC1-L3}
\end{eqnarray}

\subsection{Comparison with other BIC extensions}
{The order-constrained BIC in }\eqref{BIC1-L3} {shows some similarities with the BIC extensions proposed by Romeijn et al. (2012) and Morey \& Wagenmakers (2014). In the proposal of Romeijn et al., the prior can be chosen by users allowing a subjective quantification of the relative size of the constrained space. Although this may be useful in certain situations, the BIC is typically used in an automatic fashion, and thus it may be preferable to also let the prior probability be based on a default prior. The advantage of using the local unit information prior for this purpose is that it results in a reasonable default measure for the relative size of an order-constrained parameter space because the prior is centered on the boundary of the constrained space (unlike the (nonlocal) unit information prior). For example, when considering a univariate one-sided constraint, $\theta<0$, the prior probability based on the local unit information prior will be $\frac{1}{2}$, which seems reasonable because half of the unconstrained space of $\theta$ is covered by the one-sided constraint.}

{Furthermore Romeijn et al. fix the posterior probability that the order constraints hold to 1 in the case the MLE is in agreement with the constraints, and 0 elsewhere. This additional approximation step follows directly from large sample theory: When the sample size goes to infinity the posterior probability converges to 1 if the true parameter value is an interior point of the order-constrained subspace, and 0 if it is an interior point of the complement of this subspace. Thus, for extremely large samples, the prior-adapted BIC of Romeijn et al. may perform similar to the order-constrained BIC in }\eqref{BIC1-L3}. {For modestly sized samples however, or in the case of small effects (as is typical in social research), fixing the posterior probability to either 1 or 0 may result in crude approximations of the posterior probability. As will be shown in the empirical application in Section 6.1 for example, the posterior probabilities that two competing sets of order constraints hold under an unconstrained model are equal to .50 and .18. Setting these probabilities to 1 and 0, respectively, would result in an unnecessarily crude estimate of the marginal likelihood. Instead we recommend using the actual posterior probability that the order constraints hold based on the unconstrained approximated posterior (the third term in }\eqref{BIC1-L3}).

{In the proposal of Morey \& Wagenmakers (2014) the prior probability that a specific ordering of $d$ parameters hold, e.g., $\theta_1<\ldots<\theta_d$, is set to $1/d!$. This probability is thus based on the assumption that each ordering is equally likely a priori, similar as the priors proposed by} \cite{Mulder:2010} and \cite{Klugkist:2005} {when using Bayes factors. This probability however only holds for specific covariance structures, such as a diagonal covariance structure. The prior probability may not be invariant for reparameterizations of the model }\citep[see also][]{Mulder:2014a}. {For example, if we would define $\xi_{d'}=\theta_{d'}-\theta_{d'-1}$, for $d'=2,\ldots,d$, and $\xi_1=\theta_1$, the above order constraints would be equivalent to the one-sided constraints $(\xi_1,\ldots,\xi_{d-1})>\textbf{0}$. If one would use a prior diagonal covariance structure for $\bm\xi$ and zero means, the prior probability would be equal to $1/2^{d-1}$. This may be very different from $1/d!$, resulting in a serious violation of invariance to reparamaterizations. The prior probability based on the local unit information (the fourth term in }\eqref{BIC1-L3}) {on the other hand would be invariant for such reparameterizations as the prior covariance structure automatically transforms along with the reparameterization.}

\section{Numerical analyses}
The behavior of approximated Bayes factors based on the unit information prior and the local unit information prior will be investigated in a numerical example of the linear regression model, $y_i=\theta_0+\theta_1x_{i1}+\theta_2x_{i2}+\epsilon_i$, with $\epsilon_i\sim N(0,\sigma^2)$, for $i=1,\ldots,n$, $\theta_0$ is the intercept, and $\theta_1$ and $\theta_2$ are the effects of the first and second predictor. We consider a model selection problem between an order-constrained model $M_1:\theta_2>\theta_1>0$, a null model $M_0:\theta_2=\theta_1=0$, and the complement model, $M_2:\theta_2\not\ge\theta_1\not\ge0$. To gain more insight into the behavior of the criterion as an Occam's razor, we also test the order-constrained model $M_1$ against the unconstrained model, $M_u:(\theta_1,\theta_2)\in\mathbb{R}^2$.

\subsection{Statistical evidence for order-constrained models}
To understand better how the approximated Bayes factors quantify statistical evidence for order constrained models,
we computed the approximated Bayes factors for data with $(\hat{\theta}_1,\hat{\theta}_2)=(a,2a)$, for $a\in(-1.5,1.5)$, while fixing $\hat{\theta}_0=0$, $\hat{\sigma}^2=1$, $n=20$, and $\textbf{X}'\textbf{X}=[n~0~0;~0~n~n/2;0~n/2~n]$ (the exact choice of these fixed values did not qualitatively affect the results). Thus there is evidence for $M_1$, $M_0$, and $M_2$ when $a>0$, $a=0$, and $a<0$, respectively.

\begin{figure}[t]
\centering
\makebox{\includegraphics[width=13.8cm]{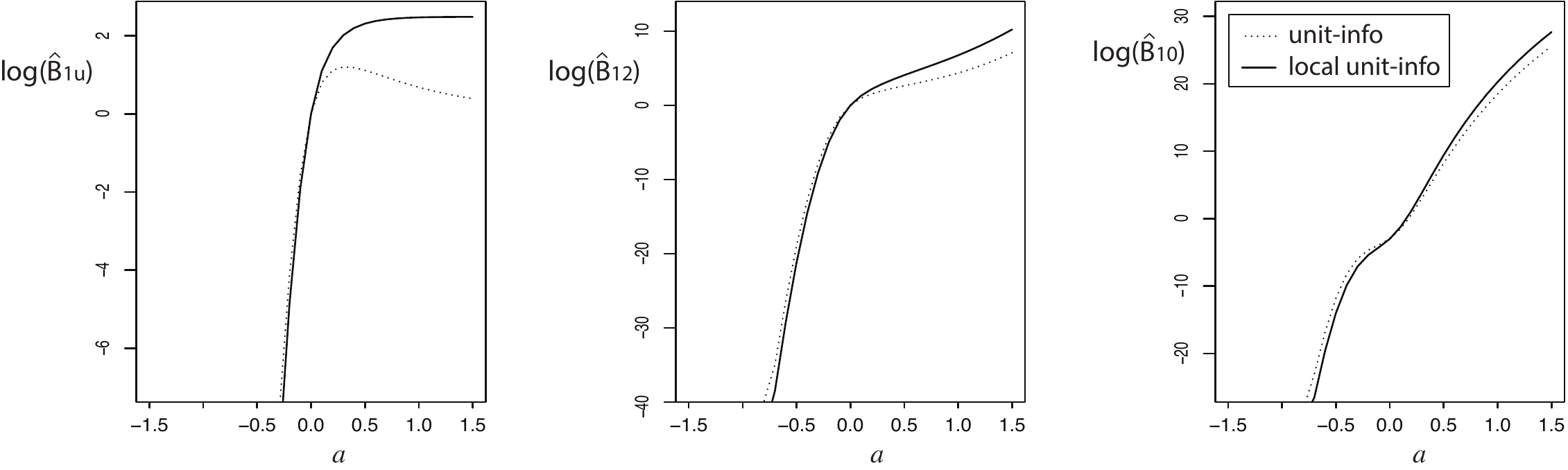}}
\caption{The logarithm of the approximated Bayes factors based on the unit information prior $\log(\hat{B}^{UI})$ (dotted line), and the local unit information prior $\log(\hat{B}^{L})$ (solid line) of $M_1:\theta_2>\theta_1>0$ versus $M_u:(\theta_1,\theta_2)\in\mathbb{R}^2$ (left panel), of $M_1$ versus the complement model $M_2$ (middle panel), and of $M_1$ versus $M_0:\theta_2=\theta_1=0$ (right panel). The criteria are plotted for $n=20$ as a function of $a$, where $(\hat{\theta}_1,\hat{\theta}_2)=(a,2a)$.}
\label{fig_sim_test1}
\end{figure}

The logarithm of the approximated Bayes factors can be found in Figure \ref{fig_sim_test1}. Based on the the approximated Bayes factors of $M_1$ versus $M_u$ (left panel) we can see that the evidence based on the unit information prior (dotted line) for $M_1$ against $M_u$ starts to decrease for larger effects (for approximately $a=.3$ and larger), which seems counterintuitive. Eventually the weight of evidence (i.e., the log Bayes factor) converges to 0. Thus, in the case of overwhelming evidence for an order constrained model, we obtain equal evidence for an order-constrained model $M_1$ that is fully supported by the data and the `larger' unconstrained model $M_u$ when using the unit-information prior, even though $M_1$ is a simpler model. This suggests that the approximated Bayes faction based on the unit information prior does not work as an Occam's razor when evaluating order constrained models. 

The evidence for $M_1$ against $M_u$ based on the local unit-information prior (solid line) on the other hand increases as a function of $a$. Eventually the weight of evidence converges to the reciprocal of the prior probability of $\theta_2>\theta_1>0$ under the unconstrained model $M_u$ which is strictly larger than 0. Furthermore the local unit-information prior results in more evidence for a model that is supported by the data in comparison to the unit-information prior when comparing model $M_1$ versus model $M_2$ (Figure \ref{fig_sim_test1}, middle panel) and model $M_1$ versus $M_0$ (Figure \ref{fig_sim_test1}, right panel). Based on these considerations we can conclude that the order-constrained BIC based on the local unit-information prior better balances fit and complexity when evaluating order-constrained models than the order-constrained BIC based on the unit-information prior.

\subsection{Error probabilities}
Next we investigate the probabilities of selecting the true data generating model when including order constraints in the alternative model or not. First we consider testing the null model, $M_0:\theta_1=\theta_2=0$, against an unconstrained alternative, $M_u:\bm\theta\in\mathbb{R}^2$, using the ordinary BIC. Second we consider testing the null model $M_0:\theta_1=\theta_2=0$ versus $M_1:\theta_2>\theta_1>0$ against two order-constrained alternative, namely $M_2:\theta_2\not\ge\theta_1\not \ge 0$, using the two order-constrained BICs. Note that the BIC for $M_0$, with no inequality constraints, is the same in both tests. Further note that because the second test contains three models instead of two, the error probabilities in the second selection problem will be a bit larger when $M_0$ is true, as a result of the design. The true effects will be set to $(\theta_1,\theta_2)=(a,2a)$, for $a=0$, so that $M_0$ is true, and $a=.1$, .2, and .4, so that $M_u$ ($M_1$) is true in the first (second) test.

\begin{figure}[t]
\centering
\makebox{\includegraphics[width=11.5cm]{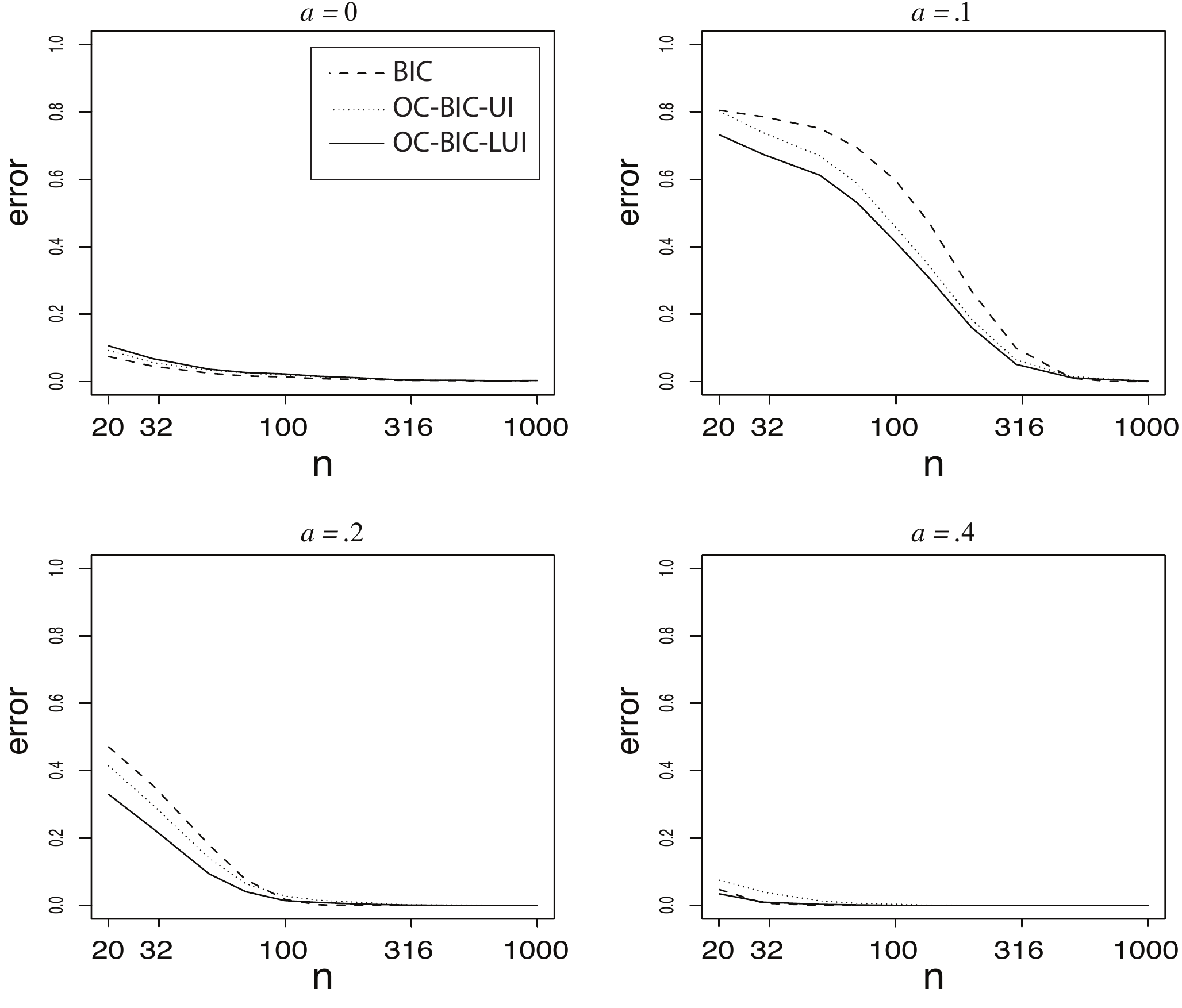}}
\caption{Probability of selecting the wrong model when using the ordinary BIC for testing $M_0:\theta_1=\theta_2=0$ against $M_u:\bm\theta\in\mathbb{R}^2$ (dashed line), and the two order-constrained BICs when testing $M_0:\theta_1=\theta_2=0$, $M_1:\theta_2>\theta_1>0$, and $M_2:\theta_2\not\ge\theta_1\not \ge 0$ (dotted and solid line for the non-local and local unit-information prior, respectively) for true effects of $(\theta_1,\theta_2)=(a,2a)$, for $a=0$, $.1$, .2, and .4. The sample size on the x-axis is on a logarithmic scale.}
\label{fig_error1}
\end{figure}

Figure \ref{fig_error1} displays the error probabilities as a function of the sample size (on a log-scale). All criteria show consistent behavior in the sense that the error probabilities go to zero as the sample size grows. Furthermore we see that when $M_0$ is true (upper left panel), the error probabilities are very similar and the ordinary BIC in test 1 results in the smallest errors. In the case of a true effect in the direct of the order constraints of $M_1$ we see that the order-constrained BIC based on the local unit-information prior results in a considerably smaller errors than the other criteria. 

The error probabilities of the order-constrained BIC based on the local unit-information prior were only slightly larger in the case of a non-zero effect. 
This is partly a consequence of the design of the test having three instead of two models under investigation. 
We conclude that overall the order-constrained BIC based on the local unit-information prior performs best in terms of error probabilities.

\subsection{Approximation errors of the order-constrained BICs}
Finally we investigated the relative approximation errors of the order-constrained BICs by comparing them to nonapproximated counterparts, e.g., $\frac{\log B_{12} - \log \hat{B}_{12}}{\log B_{12}}$ for model $M_1$ against $M_2$. The approximation errors were investigated when the order-constrained model is supported by the data, namely when testing $M_1:\theta_2>\theta_1>0$ versus $M_0:\theta_1=\theta_2=0$, with $(\hat{\theta}_1,\hat{\theta}_2)=(.5,1)$, and when the order-constrained is not supported by the data, namely when testing $M_1:\theta_2>\theta_1>0$ against its complement $M_2:\theta_1\not >\theta_2\not >0$, with $(\hat{\theta}_1,\hat{\theta}_2)=(-.5,-1)$, while increasing the sample size.

\begin{figure}[t]
\centering
\makebox{\includegraphics[width=11.5cm]{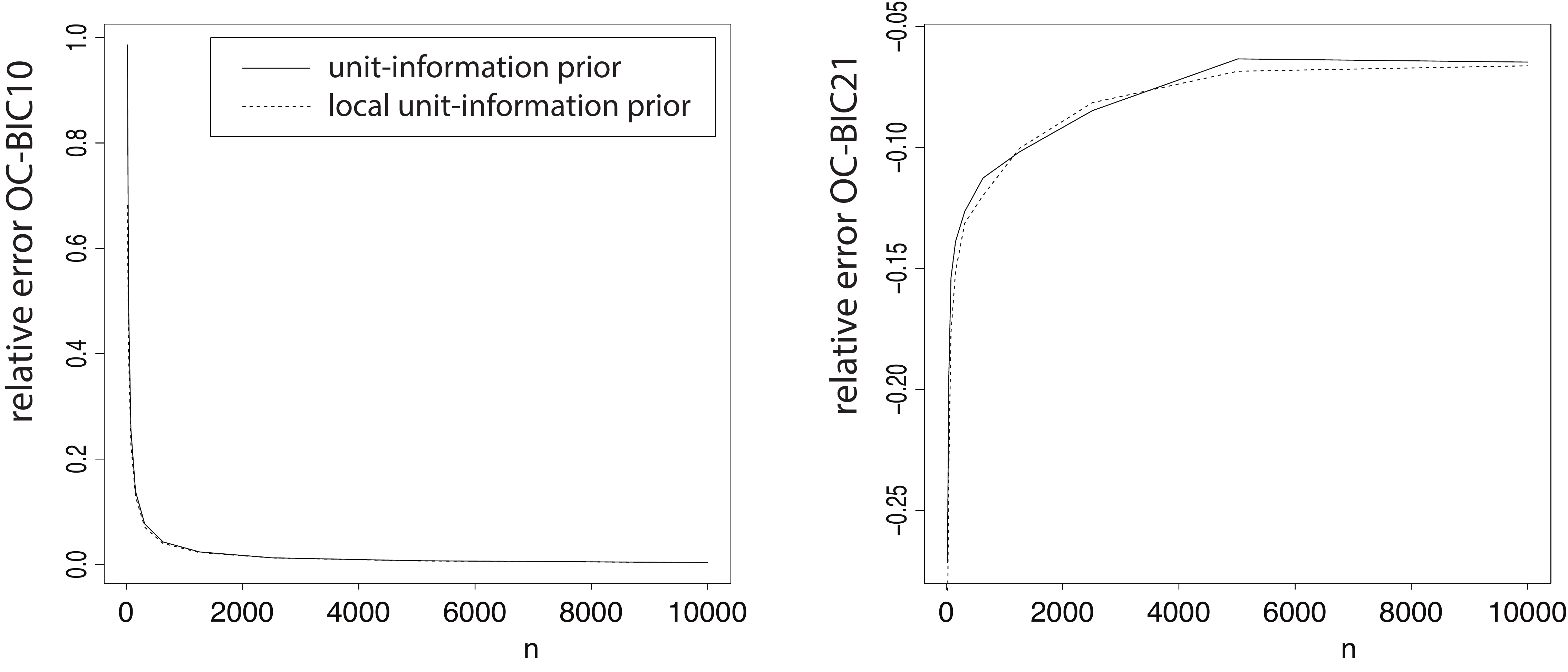}}
\caption{Left panel: Relative approximation error of the order-constrained BIC of $M_1:\theta_2>\theta_1>0$ versus $M_0:\theta_1=\theta_2=0$ when $(\hat{\theta}_1,\hat{\theta}_2)=(.5,1)$. Right panel: Relative approximation error of the order-constrained BIC of $M_2:\theta_1\not >\theta_2\not >0$ versus $M_1:\theta_2>\theta_1>0$ when $(\hat{\theta}_1,\hat{\theta}_2)=(-.5,-1)$.}
\label{fig_approxerror}
\end{figure}

The results can be found in Figure \ref{fig_approxerror}. As can be seen from the left panel, the relative error goes to 0 fast when the effects are in agreement with the order constraints of model $M_1$. When the effect are not in agreement with the constraints (right panel), we see that the relative error does not go to zero. This is a consequence of the somewhat crude approximation we already observed in Figure \ref{figApprox} (red line). The approximation error however is not large enough to be a serious practical problem. Other settings resulted in qualitatively similar results.

\section{Software}
The {\tt R}-package `{\tt BICpack}' was developed for evaluating order-constrained models using the order-constrained BIC based on the local unit-information prior. The {\tt R}-functions can be downloaded from \textsf{www.github.com/jomulder/BICpack}\footnote{Run `{\tt devtools::install\_github("jomulder/BICpack")}' in {\tt R} to install the {\tt BICpack} package.}, or from CRAN in the near future. The order-constrained BIC based on the truncated unit-information prior was not considered because of its poorer performance we observed in the numerical simulations. The package make use of the {\tt mvtnorm}-package \citep{Genz:2016} for computing the probabilities in \eqref{BIC1-L2}. The key function is `{\tt bic\_oc}', which can be used for computing the order-constrained BIC for various statistical models, including generalized linear models and survival models. As input the function needs a fitted model object (e.g., a fitted {\tt glm}-object or {\tt coxph}-object), a character string denoting the order constraints on certain parameters, and a boolean argument denoting whether the order-constrained subspace or its complement is considered (default is the order-constrained subspace).

For example, in the case of a regression model with three predictors, say, {\tt X1}, {\tt X2}, and {\tt X3}, on an outcome variable {\tt y}, and it is expected that {\tt X1} has the largest effect on the outcome variable, followed by {\tt X2}, and {\tt X3} is expected to have the smallest effect, and all effects are expected to be positive, the order-constrained BIC can be computed by executing the following lines
\begin{verbatim}
fit1 <- glm(y ~ X1 + X2 + X3, data)
bic_oc(fit1, "0 < X1 < X2 < X3")
\end{verbatim}

The use of the function will be illustrated in two empirical applications in the next section.

\section{Empirical applications revisited}
The models from the applications in Section 2 are evaluated using the order-constrained BIC based on the local unit information prior using the {\tt R}-package {\tt BICpack}. For the empirical analyses presented in this section, only the European Values Study data of Germany are considered. 

\subsection{Application 1: Assessing the importance of different dimensions of socioeconomic status}
Model \eqref{model1} can be fitted in {\tt R} using the {\tt lm}-function:
\begin{verbatim}
lm1 <- lm(atti_immi ~ class + education + income + gender,
          data=EVS_Germany)
\end{verbatim}
The estimated coefficients of interest were $(\beta_{\text{class}},\beta_{\text{education}},\beta_{\text{income}})=(.312,.250,.041)$, with standard errors .067, .075, .072, respectively.

The order-constrained BIC for model $M_1:\theta_{\text{class}}>\theta_{\text{education}}>\theta_{\text{income}}>0$ in \eqref{App1} can then be computed using the new {\tt bic\_oc}-function:
\begin{verbatim}
bic_oc(lm1, "class > education > income > 0")
\end{verbatim}
This resulted in a BIC of 3918.46. The function also provides the posterior probability that the constraints hold under the unconstrained model, which was equal to .50. Next the BIC for model $M_2:\theta_{\text{education}}>(\theta_{\text{class}},\theta_{\text{income}})>0$ is computed using
the command
\begin{verbatim}
bic_oc(lm1, "education > (class, income) > 0")
\end{verbatim}
The resulting OC-BIC was 3921.98. For this set of constraints, the posterior probability under the unconstrained model equaled .18.

The BIC for model $M_3:\theta_{\text{class}}=\theta_{\text{education}}=\theta_{\text{income}}>0$ is computed. First the model is fitted with the equality constraints on the effect but without the inequality constraint. Because the effects of social class, education, and income are equal under $M_3$, the regression model in \eqref{model1} becomes
\begin{eqnarray*}
\text{atti}\_\text{immi} &=& \theta_0 ~+ ~(\text{class}~+ ~\text{education}~+~\text{income})\times \theta_{\text{class.educ.income}}+\text{error}
\end{eqnarray*}
where $\theta_{\text{class.educ.income}}$ denotes the equal effect of social class, educational level, and income on attitude towards immigrants. Thus, this model can be fitted by including the sum of the class, education, and income as a linear predictor: 
\begin{verbatim}
EVS_Germany$class.educ.income <- EVS_Germany$class +
   EVS_Germany$education + EVS_Germany$income
lm2 <- lm(atti_immi ~ class.educ.income + gender, data=EVS_Germany)
\end{verbatim}
The order-constrained BIC can then be computed based on the resulting fitted model:
\begin{verbatim}
bic_oc(lm2, "class.educ.income > 0")
\end{verbatim}
This resulted in a BIC of 3917.84.

Finally, to compute the BIC of the complement model $M_4:\text{``neither $M_1$, $M_2$, nor $M_3$,''}$, first note that the marginal likelihood of the union of $M_1$, $M_2$, and $M_3$ would be the same as the marginal likelihood of the union of only $M_1$ and $M_2$ because $M_3$ has zero probability due to the presence of the equality constraints of $M_3$. Thus, we need to compute the marginal likelihood of the complement model of the joint of models $M_1$ and $M_2$. First we combine the two sets of order constraints in one vector, and then compute the order-constrained BIC using the new function:
\begin{verbatim}
constraints_M4 <- c("class > education > income > 0",
                     "education > (class, income) > 0") 
bic_oc(lm1, constraints_M4, complement = TRUE)
\end{verbatim}
This resulted in a BIC of 3926.13. The BIC values are summarized in Table \ref{tab1}. From these values we can conclude that model $M_3$ receives most support but the evidence is negligible in comparison to the evidence for the order-constrained model $M_1$, given the BIC-difference of .62\footnote{Typically a BIC-difference of 10 points is needed in order to rule out a model.}. The evidence for $M_2$ and $M_3$ is considerably lower than for $M_1$ and $M_3$.

For interpretation purposes it can be useful to translate the BICs to posterior model probabilities. A posterior model probability quantifies the probability of the data having been generated by one of the models considered, after observing the data given certain prior model probabilities. This probability is conditional on the data having been generated by one of the models considered.

In this application we assume equal prior probabilities for the models. The posterior model probabilities can be computed from the BIC-values using the `{\tt postprob}' function in {\tt BICpack}. The posterior probabilities together with the BICs can be found in Table \ref{tab1}. Hence the posterior probability for model $M_3$, which assumes equal and positive effects of social class, education, and income on attitude towards immigrants, is largest with 53.3\%. The posterior probability of $M_1$, which assumed ordered positive effects of social class, education, and income based on the Ethnic Competition Theory, is only slightly smaller with 39.1\%. There is not much evidence for either $M_2$ or $M_2$, given  their posterior probabilities of 6.7\% and .8\%, respectively. There is thus considerable model uncertainty, and more data would be needed to choose a single best model.

\begin{table}[t]
\begin{center}\label{simresults}
\caption{Order-constrained BICs and posterior model probabilities for the competing models in Application 1.}
\begin{tabular}{lcc}
\hline
 & OC-BIC$^*$ & $P(M_t|\textbf{D})$\\
\hline 
$M_1:\theta_{\text{class}}>\theta_{\text{education}}>\theta_{\text{income}}>0$ & 3918.46 & 0.391\\
$M_2:\theta_{\text{education}}>(\theta_{\text{class}},\theta_{\text{income}})>0$ & 3921.98 & 0.067\\
$M_3:\theta_{\text{class}}=\theta_{\text{education}}=\theta_{\text{income}}>0$ & 3917.84 & 0.533\\
$M_4:\text{``neither $M_1$, $M_2$, nor $M_3$,''}$ & 3926.13 & 0.008\\
\hline
\end{tabular}\label{tab1}
\end{center}
\end{table}

\subsection{Application 2: The importance of postmaterialism for young, middle and old generations}
Because the outcome variable `postmaterialism' has an ordinal measurement level with three categories (`low', `medium', and `high'), an ordinal regression model can be fitted using the {\tt polr}-function of the {\tt MASS}-package. Thus, the ordinal variable `postmaterialism' is regressed to the ordinal predictor `generation' with categories `young', `middle', and `old' while controlling for `gender', `income', and `education':
\begin{verbatim}
fit3 <- polr(postmaterial ~ generation + gender + income +
          education, data=EVS_Germany, Hess=TRUE)
\end{verbatim}
In the fitted model the `young' generation is the reference group and dummy variables are created for the `middle' and `old' generation. These variables are called `{\tt generationmiddle}' and `{\tt generationold}' in the fitted {\tt polr}-object. The estimated effects under this model were equal to $(\hat{\theta}_{\text{generationmiddle}},\hat{\theta}_{\text{generationold}})=(-.444,-.848)$, having standard errors of .154 and .150, respectively. 

Thus, the order-constrained BIC of model $M_1:\theta_{\text{generationold}}<\theta_{\text{generationmiddle}}<0$, representing the Generational Replacement Theory, can be computed by the command
\begin{verbatim}
bic_oc(fit3, "generationold < generationmiddle < 0")
\end{verbatim}
The resulting BIC equaled 3154.82.

Next, the BIC of the null model $M_0:\theta_{\text{generationold}}=\theta_{\text{generationmiddle}}=0$ is computed with no generation effect. Because this model does not contain any order constraints, we can simply compute an ordinary BIC. This can also be done using the {\tt bic\_oc}-function by omitting any order constraints:
\begin{verbatim}
fit4 <- polr(postmaterial ~ 1 + gender + income + 
          education, data=EVS_Germany, Hess=TRUE)
bic_oc(fit4)
\end{verbatim}
The resulting BIC was equal to 3177.69.

Finally the BIC of the complement model was computed. Similarly to the previous example, this can be done as follows
\begin{verbatim}
bic_oc(fit3, "generationold < generationmiddle < 0", complement=TRUE)
\end{verbatim}
This resulted in a BIC of 3170.15. The BICs and respective posterior model probabilities can be found in Table \ref{tab2}. Clearly, there is overwhelming evidence for $M_1$ which implies that postmaterialism has increased for younger generations.

\begin{table}[t]
\begin{center}\label{simresults}
\caption{Order-constrained BICs and posterior model probabilities for the competing models in Application 2.}
\begin{tabular}{lcc}
\hline
 & OC-BIC$^*$ & $P(M_t|\textbf{D})$\\
\hline 
$M_0:\theta_{\text{old}}=\theta_{\text{middle}}=0$ & 3177.69 & 0.00\\
$M_1:\theta_{\text{old}}<\theta_{\text{middle}}<0$ & 3154.82 & 1.00\\
$M_2:\text{``neither $M_0$, nor $M_1$,''}$ & 3170.15 & 0.00\\
\hline
\end{tabular}\label{tab2}
\end{center}
\end{table}

Finally we show that the inclusion of order constraints in the alternative model results in more evidence against a null model if the order constraints are supported by the data. First note that the BIC for the order-constrained model, $M_1$, against the null model, $M_0$, equals $\text{BIC}(M_{1},M_0)=\text{BIC}(M_{1})-\text{BIC}(M_{0})=3154.82-3177.69=-22.87$. The BIC for an unconstrained alternative model, $M_3:\theta_{\text{generationold}}\not=\theta_{\text{generationmiddle}}\not=0$, against the null model equals $\text{BIC}(M_{3},M_0)=\text{BIC}(M_{3})-\text{BIC}(M_{0})=3158.18-3177.69=-19.51$\footnote{The BIC for $M_3$ can be obtained by running `{\tt bic\_oc(fit3)}'.}. Hence, the inclusion of order constraints results in a substantial increase of the evidence against the null model in the case the order constraints are supported by the data. We also get a more informative answer about how the effects are related to each other in the case there is evidence against the null model than when testing the null against an unconstrained alternative.

\section{Discussion}
In this paper we presented two extensions of the BIC for evaluating models with order constraints on certain parameters of interest. In the first extension a truncated unit-information prior was considered under the order-constrained model and in the second extension a truncated local unit-information prior was considered. Theoretical considerations and numerical analyses revealed that the local unit-information prior resulted in better model selection behavior than the non-local unit information prior for order-constrained model selection.

The new order-constrained BIC based on the local unit-information prior can easily be computed using the new {\tt R}-package `{\tt BICpack}`. This will allow researchers to test multiple social theories that can be translated into conflicting sets of equality and order constraints on the parameters of interest. The methodology can also be used for testing directed effects of ordinal predictors, as these expectations can be translated into order-constrained models in a natural manner.

\section*{Acknowledgements}
The authors thank Anton Olsson Collentine for helping with the R-code for reading order constraints, and Tim Reeskens and John Gelissen for useful insights on the empirical applications from the European Values Study. Mulder's research was supported by a NWO Vidi grant (452-17-006). Raftery's research was supported by NIH grants R01 HD054511 and R01 HD 070936,
and by the Center for Advanced Study in the Behavioral Sciences at Stanford
University.

\bibliographystyle{apacite}
\bibliography{MulderRaftery_BIC-for-order-constraints_v14}

\end{document}